\documentclass[aps,prb,twocolumn,groupedaddress,showpacs]{revtex4-1}

\usepackage[dvips]{graphicx}
\usepackage{dcolumn}
\usepackage{bm}
\usepackage{amssymb}
\usepackage{epstopdf}

\begin{document}

\title{Layered pnictide-oxide Na$_2$Ti$_2$Pn$_2$O (Pn=As, Sb): a paradigm for spin density waves}

\author{Xun-Wang Yan$^{1,2}$}
\author{Zhong-Yi Lu$^{1}$}\email{zlu@ruc.edu.cn}

\date{\today}

\affiliation{$^{1}$Department of Physics, Renmin University of
China, Beijing 100872, China}

\affiliation{$^{2}$School of Physics and Electrical Engineering, Anyang Normal University,
Anyang 455002, China}

\begin{abstract}

From the first-principles calculations, we have studied the electronic and magnetic structures of compound Na$_2$Ti$_2$Pn$_2$O (Pn = As or Sb). We find that in the ground state Na$_2$Ti$_2$As$_2$O is a blocked checkerboard antiferromagnetic semiconductor with a small band gap of about 0.15 eV, in contrast, Na$_2$Ti$_2$Sb$_2$O is a bi-collinear antiferromagnetic semimetal, both with a small moment of about 0.5$\mu_B$ around each Ti atom. We show that there is a strong Fermi surface nesting in Na$_2$Ti$_2$Pn$_2$O. And we verify that the blocked checkerboard and bi-collinear antiferromagnetic states both are the spin density waves induced by the Fermi surface nesting. A tetramer structural distortion is found in company with the formation of a blocked checkerboard antiferromagnetic order, in good agreement with the experimentally observed commensurate structural distortion but with space group symmetry retained after the anomaly happening. Further analysis and discussion in connection with experimental observations are given as well.

\end{abstract}


\maketitle

\section{Introduction}

Layered compounds of transition-metal elements frequently show interesting and novel electrical  and magnetic properties and have been studied extensively. Among them, La$_2$CuO$_4$
\cite{muller} and LaFeAsO \cite{kamihara} as the representatives are widely known because of
their superconductivity with high critical temperatures after doping or under high pressures,
in which the CuO$_2$ square planar layers and the FeAs tetrahedral layers play a substantial
role respectively. Now another class of layered compounds built from alternatively stacking
special octahedral layers M$_2$Q$_2$O (M=Fe or Ti; Q=As, Sb, S, or Se) with otherwise layers are attracting much attention.\cite{adam,mayer} Here an M$_2$Q$_2$O layer includes an
anti-CuO$_2$-type M$_2$O square planar layer with two Q atoms respectively located above and
below the center of each M$_2$O square unit. As shown in Fig. \ref{figa}, the M$_2$Q$_2$O layer is in an octahedral layer structure, bridging or a combination of the CuO$_2$ square planar
layer and the FeAs tetrahedral layer structures, which is thus expected to show distinct
electronic and magnetic properties and possibly superconductivity by doping or under pressure.

Fe$_2$Q$_2$O layer-based (Q=S or Se) oxychalcogenides were first synthesized in 1992 \cite{mayer} and recently characterized to be antiferromagnetic insulators with a magnetic moment of $4\mu_B$ on each Fe ion experimentally\cite{kabbour} and with strong correlation effects suggested theoretically.\cite{zhujianxin} To our knowledge, there has been no any structural distortion found experimentally in these compounds.

Ti$_2$Q$_2$O layer-based (Q=As or Sb) oxypnictides have been also studied for a quite long time. Compounds Na$_2$Ti$_2$Q$_2$O were first synthesized in 1990,\cite{adam} which subsequently have been intensively investigated both experimentally and theoretically.\cite{ozawa1,ozawa2,ozawa3,ozawa4} The temperature-dependent powder
neutron diffraction measurement showed that there is a structural distortion in the Ti$_2$Sb$_2$O layer at a temperature of about 120K,\cite{ozawa2,ozawa3} which corresponds well to the observed anomalies in the temperature-dependent magnetic susceptibility and
resistivity measurements.\cite{adam,ozawa1} However, no neutron scattering due to magnetic spin ordering was detected. Meanwhile, the measurement further found that the bond distance ratio of O-Ti-O/Sb-Ti-Sb increases below the transition temperature, but the crystal space group symmetry remains unchanged after the structural distortion.\cite{ozawa2,ozawa3} This is different from the tetragonal-orthorhombic structural distortions observed in the iron pnictides.\cite{cruz}

Recently another new Ti$_2$As$_2$O layer-based oxypnictide BaTi$_2$As$_2$O was successfully synthesized,\cite{chenxianhui1,chenxianhui2} which shows the similar characteristics as Na$_2$Ti$_2$Q$_2$O, namely there are anomalies observed as well in the temperature-dependent magnetic susceptibility, resistivity, and heat capacity measurements at a temperature of about 200K.\cite{chenxianhui1,chenxianhui2} Considering that the similar anomalies are observed in the iron pnictides, in which the superconductivity takes place when the anomaly is suppressed, one thus speculates that the Ti$_2$Q$_2$O layer-based oxypnictides may also be another parent compounds for the high T$_c$ superconductivity. In the iron pnictides, those anomalies are ascribed to the tetragonal-orthorhombic structural transitions accompanied by the collinear antiferromagnetic ordering.\cite{cruz} The underlying mechanism is still in debate, either the As-bridged antiferromagnetic superexchange interactions between fluctuating Fe local moments \cite{ma1,si,yildirim} or spin density wave instability induced by the Fermi surface nesting.\cite{mazin} There are now more and more evidences in favor of the fluctuating Fe local moment picture. Especially, the neutron inelastic scattering experiments have shown that the low-energy magnetic excitations can be well described by the spin waves based on the quantum Heisenberg model.\cite{dai-0,dai-1}

Likewise, the scenario in Ti$_2$Q$_2$O layer-based oxypnictides is very confusing. The experimentally observed anomalies suggest that there is an antiferromagnetic transition, but no magnetic ordering has been detected. Theoretically, no ordered magnetic state has been either found in the first-principles electronic structure calculations,\cite{pickett} which however show that there is a Fermi surface nesting. It was thus suggested that the observed anomalies would be due to charge density wave or spin density wave instability induced by such a Fermi surface nesting.\cite{pickett,fabrizia}

In order to clarify the issue, we performed the thorough first-principles electronic structure calculations upon compounds Na$_2$Ti$_2$Q$_2$O (Q=As, Sb). We find that the ground state of Na$_2$Ti$_2$As$_2$O is a blocked checkerboard antiferromagnetic semiconductor, in contrast, the ground state of Na$_2$Ti$_2$Sb$_2$O is a bi-collinear antiferromagnetic semimetal. And we verify that these two antiferromagnetic states both are indeed induced by the Fermi surface nesting rather than based on local magnetic moments.

\section{Computational details}

In our calculations the plane wave basis method was used.\cite{baroni} We used the PW91-type\cite{perdew} generalized gradient approximation for the exchange-correlation functionals. The ultrasoft pseudopotentials \cite{vanderbilt} were used to model the electron-ion interactions. After the full convergence test, the kinetic energy cut-off and the charge density cut-off of the plane wave basis were chosen to be 600eV and 4800eV, respectively. The Gaussian broadening technique was used and a mesh of $20\times 20\times 10$ k-points were sampled for the Brillouin-zone integration in nonmagnetic state. The lattice parameters with the atomic positions were fully optimized by the energy minimization.
It was found that the calculated lattice parameters have an excellent agreement with the experimental ones (less than $ 1.0 \%$).

\section{Results and analysis}

\begin{figure}
\includegraphics[width=6cm]{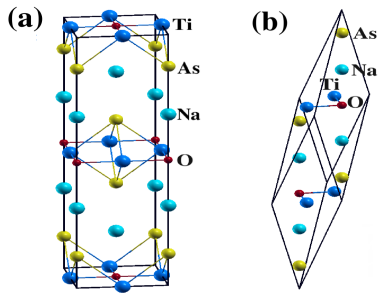}
\caption{(Color online)  Crystal structure of Na$_2$Ti$_2$As$_2$O: (a) A conventional unit cell composed of two formula cells; (b) A primitive unit cell composed of one formula cell.} \label{figa}
\end{figure}

Compound Na$_2$Ti$_2$Pn$_2$O (Pn=As or Sb) crystallizes in tetragonal (I4/mmm) symmetry and has a layered structure, as shown in Fig. \ref{figa}(a). Its conventional tetragonal unit cell
consists of two formula unit cells. However, its primitive unit cell is constructed by
considering Na$_2$Ti$_2$Pn$_2$O as a triclinic crystal, in which only one formula unit cell is
included as shown in Fig. \ref{figa}(b). On the other hand, Na$_2$Ti$_2$Pn$_2$O can be
considered geometrically with an anti-K$_2$NiF$_4$ type structure, where a Ti atom is located
between two oxygen atoms in a plane to form a square planar layer of Ti$_2$O, which is an
anti-configuration to the CuO$_2$ layer in the high $T_c$ cuperates. Meanwhile a Ti atom is
also four-coordinated by As atoms, forming a Ti$_2$As$_2$O layer where two As atoms are
respectively located above and below the center of a Ti$_2$O square unit. The crystal is formed by stacking such Ti$_2$Pn$_2$O layers in a body-centered manner along the $c$ axis with the
separation by a double-layer of Na atoms (Fig. \ref{figa}(a)).

\subsection{Nonmagnetic States}

We first studied the nonmagnetic state of compound Na$_2$Ti$_2$Pn$_2$O (Pn=As or Sb), which would describe the high temperature phase of these materials. The electronic band structure of
this nonmagnetic state also provides a reference to further studying of the possible low
temperature magnetic phases, by analyzing which we can better understand the underlying
mechanism or interactions that drive possible magnetic phase transitions and the related
structural transitions. In the calculations, we adopted a primitive unit cell including only
one formula cell (Fig. \ref{figa}(b)).

\subsubsection{Na$_2$Ti$_2$As$_2$O}

\begin{figure}
\includegraphics[width=7cm]{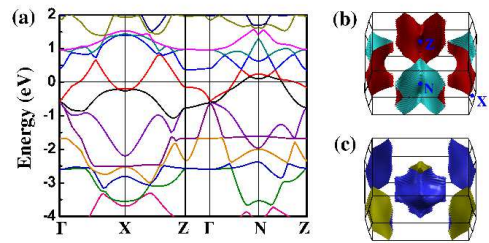}
\caption{(Color online) Electronic structure of  Na$_2$Ti$_2$As$_2$O in the nonmagnetic state: (a) Electronic Band structure; (b) Hole-type Fermi surface sheets; (c) Electron-type Fermi surface sheets. The Fermi energy is set to zero.}\label{figb}
\end{figure}

After the full structural optimization, we find the tetragonal crystal lattice parameters of $a=b=4.060$~\AA~ and $c=15.3892$~\AA~ for Na$_2$Ti$_2$As$_2$O, in good agreement with the experimental values of $a=b=4.0810$~\AA~ and $c=15.311$~\AA~ measured at a temperature of 310K. \cite{ozawa2} The calculated electronic band structure and Fermi surface in the primitive unit cell are presented in Fig. \ref{figb}. As we see, there are three Fermi surface sheets, contributed from the two bands crossing the Fermi energy level (Fig. \ref{figb}(a)). Among them, the one sheet is of the hole-type, forming an approximate square-shaped pipe centered at each $N$ point in the Brillouin zone (Fig. \ref{figb}(b)); the other two are of the electron-type, one of them is a $\Gamma$-centered box with the windows at its top and bottom, the other is a $X$-centered perfectly fluted pipe around the corner of the Brillouin zone (Fig. \ref{figb}(c)). The volumes enclosed by these Fermi surface sheets are 0.186 electrons/cell and 0.186 holes/cell respectively. The electron carrier concentration is the same as the hole carrier concentration. Both are equal to $2.94 \times 10^{21}/{cm}^3 $. Compound Na$_2$Ti$_2$As$_2$O is thus a semimetal with a low carrier concentration between the ones in normal metals and semiconductors, similar to the recent iron pnictides, for example, LaFeAsO and BaFe$_2$As$_2$.\cite{ma,ma2} The density of states at the Fermi energy is 3.84 states per eV per formula unit. The corresponding electronic specific heat coefficient is $\gamma$ = $9.05mJ/(K^2\ast mol)$ and the Pauli paramagnetic susceptibility is $\chi_p$=$1.56 \times 10^{-9} m^3/mol$ (1 mole formula unit), in reasonable agreement with the measured value of $\chi = 3.01 \times 10^{-9} m^3/mol$.\cite{ozawa3}

\begin{figure}
\includegraphics[width=7cm]{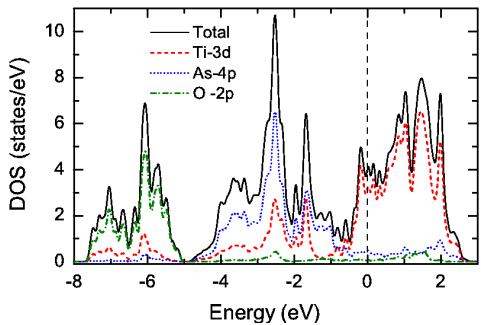}
\caption{(Color online) Total and atomic orbital-resolved partial density of states of
Na$_2$Ti$_2$As$_2$O per formula cell in the nonmagnetic state. The Fermi energy is set to
zero.}\label{figc}
\end{figure}

\begin{figure}
\includegraphics[width=7cm]{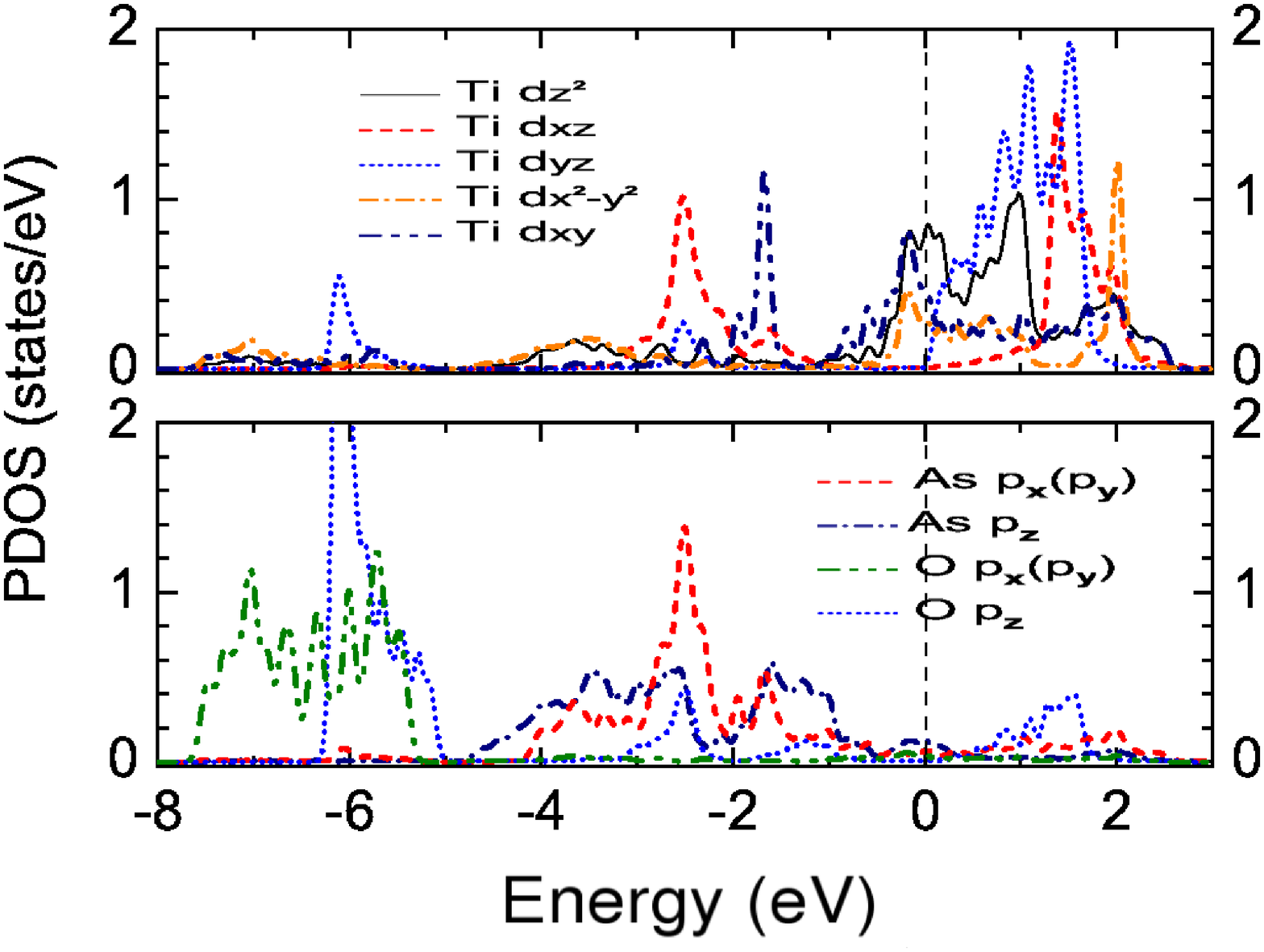}
\caption{(Color online) Density of states of Na$_2$Ti$_2$As$_2$O in the nonmagnetic state projected onto Ti-$3d$ atomic orbitals (upper panel), As-$4p$ atomic orbitals and O-$2p$ atomic orbitals (lower panel), respectively.}\label{figd}
\end{figure}

Inspection of the distribution of the density of states will help to clarify the atomic bonding character. Figure \ref{figc} plots the calculated total and atomic orbital-resolved partial
density of states of Na$_2$Ti$_2$As$_2$O in the nonmagnetic state respectively. As expected, the calculations show that most electrons of the Na $3s$-orbitals transfer into the O $2p$-orbitals so that the occupied O $2p$-orbitals are far below the Fermi energy level while the empty Na $3s$-orbitals are much above the Fermi energy level (centered at +5.0 eV). Although the density of states at the Fermi level is dominated by the Ti $3d$-orbitals, the density of states around the Fermi energy (from -3.0 to 2.0eV) consist mainly of both the Ti $3d$-orbitals and As $4p$-orbitals. Especially, the two peaks in the distribution of the As $4p$-orbitals coincide with those of the Ti $3d$-orbitals, as specifically shown in Fig. \ref{figd}, which indicates that there is the hybridization between $4p_x$ ($p_y$) orbitals and $d_{xz}$ ($d_{xy}$) orbitals, namely there is covalent bond formed between them. Figure \ref{figd} also specifies that it is the remaining three $3d$-orbitals ($d_{xy}$, $d_{x2-y2}$, and $d_{z^2}$) that dominate in the density of states at the Fermi energy.

\subsubsection{Na$_2$Ti$_2$Sb$_2$O}

\begin{figure}
\includegraphics[width=7cm]{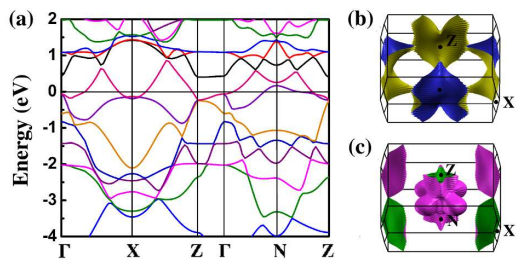}
\caption{(Color online) Electronic structure of Na$_2$Ti$_2$Sb$_2$O in the nonmagnetic state: (a) Electronic band structure; (b) Hole-type Fermi surface sheets; (c) Electron-type Fermi surface sheets. The Fermi energy is set to zero.}\label{figf}
\end{figure}

Likewise, we performed the calculations on Na$_2$Ti$_2$Sb$_2$O. The optimized tetragonal
crystal lattice parameters are found to be of $a=b=4.1369$~\AA~ and $c=16.6428$~\AA~, also in good agreement with the experimental values of $a=b=4.160$~\AA~ and $c=16.558$~\AA~ measured at a temperature of 150K.\cite{ozawa2} Figure \ref{figf} shows the electronic band structure and Fermi surface calculated in a primitive unit cell, similar to those of Na$_2$Ti$_2$As$_2$O. From the volumes enclosed by these Fermi sheets, we determine that the electron and hole concentrations are 0.140 electrons/cell and 0.140 holes/cell respectively. The corresponding electron (or hole) carrier density is about $1.97\times 10^{21}/cm^3$. The density of states at the Fermi energy is 4.40 states per eV per formula unit. The corresponding electronic specific heat coefficient and Pauli paramagnetic susceptibility are $\gamma$ = $10.37mJ/(K^2\ast mol)$ and $\chi_p$=$1.79\times 10^{-9} m^3/mol$ (1 mole formula unit), respectively. The latter is in excellent agreement with the experimental measurement on a high-quality single crystal ($\chi = 1.51\times 10^{-9} m^3 /mol$).\cite{ozawa3}

\begin{figure}
\includegraphics[width=7cm]{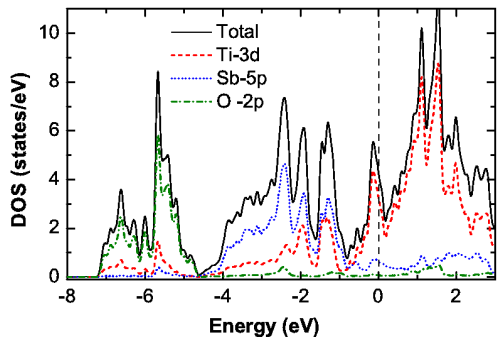}
\caption{(Color online) Total and atomic orbital-resolved partial density of states of Na$_2$Ti$_2$Sb$_2$O per formula cell in the nonmagnetic state. The Fermi energy is set to zero.}\label{figg}
\end{figure}

In Fig. \ref{figg}, we plot the calculated total and atomic orbital-resolved partial density of states for Na$_2$Ti$_2$Sb$_2$O in the nonmagnetic state, which are overall similar to the ones
of Na$_2$Ti$_2$As$_2$O (Fig. \ref{figc}) except that the Sb $5p$-orbitals have more
contribution to the density of states at the Fermi energy than the As $4p$-orbitals do. This is easily understandable because the Sb $5p$-orbitals are more diffusive and extended than the As
$4p$-orbitals. Nevertheless, the atomic bonding character of Na$_2$Ti$_2$Sb$_2$O is basically
the same as the one of Na$_2$Ti$_2$As$_2$O, i.e. O and Na atoms are in ionic state while there
are covalent bonds formed between Ti and Sb atoms.

\subsubsection{Fermi Surface Nesting}

\begin{figure}
\includegraphics[width=7.5cm]{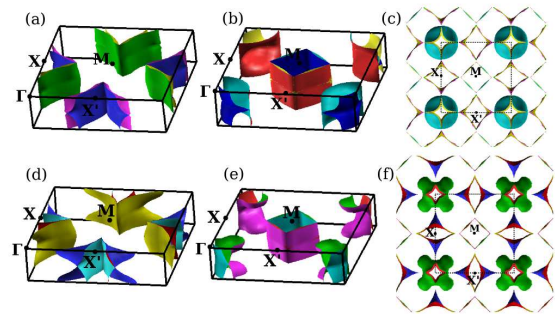}
\caption{(Color online) Na$_2$Ti$_2$As$_2$O in the nonmagnetic state: (a) and (b) are respectively the hole-type and electron-type Fermi surface sheets, and (c) is the top view of the whole Fermi surface in the Brillouin Zone denoted by the dotted square;  Na$_2$Ti$_2$Sb$_2$O in the nonmagnetic state: (d) and (e) are respectively the  hole-type and electron-type Fermi surface sheets, and (f) is the top view of the whole Fermi surface in the Brillouin Zone denoted by the dotted square.}\label{fige}
\end{figure}

In order to well analyze the Fermi surface symmetry of compounds Na$_2$Ti$_2$Pn$_2$O (Pn=As, Sb) in connection with the crystal symmetry, we replot the calculated Fermi surfaces in the tetragonal conventional Brillouin zone as shown in Fig. \ref{fige}, which clearly reflects the two dimensional character of the electronic structures by the nearly $k_z$-dispersionless Fermi surface sheets, being consistent with the layered crystal structure. As we see, the Fermi surfaces in these two compounds are very similar in shape to each other. There are now six Fermi surface sheets for each Fermi surface, among which there are the two nearly degenerate square-box-like Fermi-surface sheets centered around X contributed from the hole-type bands in conjugation with another two similar Fermi surface sheets centered around M but from the electron-type bands. Obviously these two sets of Fermi surface sheets between X and M are strongly nested with each other. The corresponding nesting vector is $(\frac{\pi}{a},0,0)$ or $(0,\frac{\pi}{a},0)$. The early studies reported in Refs. \onlinecite{pickett} and \onlinecite{fabrizia} first suggested that such a Fermi surface nesting in Na$_2$Ti$_2$Sb$_2$O would likely lead to a charge- or spin-density wave instability in association with the observed anomalies in the temperature-dependent susceptibility and resistivity.\cite{pickett,fabrizia,ozawa2}

\subsection{Magnetic States}

The powder neutron diffractions have not yet detected any magnetic spin ordering in compounds Na$_2$Ti$_2$Pn$_2$O (Pn=As, Sb) even though the observed anomalous transitions in the temperature-dependent magnetic susceptibility and electric resistivity strongly suggested an antiferromagnetic transition in the compounds. Nevertheless the neutron diffractions indeed found a commensurate structural distortion but without the space group symmetry breakdown.

Right now we still cannot exclude any weak magnetic ordering just based on the powder neutron
diffraction because of its low resolution. The issue that we meet here is still whether or not there exists a magnetic long-range order related to the structural distortion, based on either local moments formed around Ti atoms or spin-density wave induced by the Fermi surface nesting; Otherwise, whether or not there exists a spin-unpolarized charge-density wave related to the structural distortion, induced still by the Fermi surface nesting. In order to clarify this issue, we have carried out the systematical and extensive calculations on compounds Na$_2$Ti$_2$Pn$_2$O by constructing a wide variety of unit cells and magnetic orders.

We first performed the spin-unpolarized calculations with different unit cells and possible distortions, especially by elaborately constructing new unit cells in consideration of the nesting vector between X and M points in the Brillouin zone. It turns out that all the distortions and structures different from the original one are unstable. We thus exclude spin-unpolarized charge-density waves in compounds Na$_2$Ti$_2$Pn$_2$O.

We then investigated whether or not any magnetic order exists, based on local moments around Ti atoms. If there are local moments formed around Ti atoms, the exchange interactions
between these local moments would be either Pn-bridged or O-bridged superexchange interaction
with the direct exchange interaction between the nearest neighbor Ti atoms. There would be a
variety of stable or metastable magnetic orders resulting from the competition among these
exchange interactions, such as the ferromagnetic, checkerboard antiferromagnetic, and collinear antiferromagnetic orders, all of which were found to be stable or metatable in Fe-based superconducting parent compounds.\cite{ma,ma1,ma2,ma3} Our spin-polarized calculations show that none of these magnetic orders is stable or metastable in compounds Na$_2$Ti$_2$Pn$_2$O, similar to the previous calculations that show no magnetic ordering found in Na$_2$Ti$_2$Sb$_2$O.\cite{pickett} We thus exclude magnetic orders based on local moments.

To the end, we came to study possible magnetic orders based on spin-density waves induced by the Fermi surface nesting.

\subsubsection{Blocked checkerboard antiferromagnetic order and bi-collinear antiferromagnetic order}

As presented above, there are two independent and orthogonal but equivalent nesting vectors $\vec{Q_1}=(\frac{\pi}{a},0,0)$ and $\vec{Q_2}=(0,\frac{\pi}{a},0)$, respectively. A possible spin density wave will oscillate as $\vec{M}\cos(\vec{Q}\cdot\vec{R})$ on the Ti-Ti square lattice ($\vec{R}$ being a lattice site vector and $\vec{M}$ magnetization vector), with a wave vector $\vec{Q}$ equal to a commensurate linear combination of $\vec{Q_1}$ and $\vec{Q_2}$, namely $\vec{Q}=n\vec{Q_1}+m\vec{Q_2}$ ($n$ and $m$ being integers). We elaborately constructed several kinds of magnetic orders with magnetic unit cells based on such wave vectors $\vec{Q}$. Through the calculations, we eventually find two stable peculiar magnetic orders with wave vectors $\vec{Q_1}+\vec{Q_2}$ and $\vec{Q_1}$ (or $\vec{Q_2}$) respectively, called as blocked checkerboard antiferromagnetic order and bi-collinear antiferromagnetic order respectively.

\begin{figure}
\includegraphics[width=7.5cm]{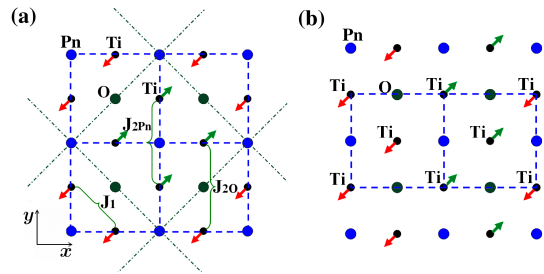}
\caption{(Color online) Schematic top view of the Ti$_2$Pn$_2$O (Pn=As, Sb) layer. The large blue circles represent the projected sites of Pn atom into the Ti$_2$O plane, the dark olive circles represent the O atoms, the small black circles with arrows represent Ti atoms and the green and red arrows indicate the two opposite directions of Ti moments. The blocked checkerboard antiferromagnetic order (a) and the bi-collinear antiferromagnetic order (b) are shown. Here a small blue dashed square represents an $a \times a$ unit in (a) and (b). The large blue dashed square $2a \times 2a$ unit in (a), and the blue dashed rectangle $2a \times a$ unit in (b) correspond to a magnetic unit cell in the blocked checkerboard antiferromagnetic order and bi-collinear antiferromagnetic order respectively. For the blocked checkerboard antiferromagnetic order, the Ti$_2$Pn$_2$O layer is divided into square blocks denoted by gray dash-dotted lines. The four Ti moments in a block are in the same direction and those between two nearest neighbor blocks have opposite directions.}
\label{fig8-two-order}
\end{figure}

Figure \ref{fig8-two-order}(a) schematically shows the blocked checkerboard antiferromagnetic order, in which the Ti$_2$Pn$_2$O (Pn=As, Sb) layer is divided into $\sqrt{2}\times\sqrt{2}$-squares represented by the dash-dotted thin lines. Each square is a magnetic block with the moments of four Ti atoms being parallel. Between the nearest neighbor blocks, the block moments are in anti-parallel. But the magnetic unit cell is a $2a\times 2a$ square. It should be emphasized that in each magnetic block the four Ti moments are centered around a Pn atom rather than O atom. For the latter case, the calculations show that it is unstable or energetically unfavorable. Thus in the blocked checkerboard anitferromagnetic order any pair of the next nearest neighbor Ti moments bridged by an O atom is always in anti-parallel. This is also consistent with the O-bridged superexchange effect being more stronger than the Pn-bridged superexchange effect, even though this magnetic order is induced by the Fermi surface nesting.

In Fig. \ref{fig8-two-order}(b), we schematically show the bi-collinear antiferromagnetic order, in which the Ti moments align in parallel along a diagonal direction and in anti-parallel along the other diagonal direction on the Ti-Ti square lattice. In other words, if the Ti-Ti square lattice is divided into two square sublattices $A$ and $B$, the Ti moments on each sublattice take their own collinear antiferromagnetic order. Here the magnetic unit cell is a $2a\times a$ rectangle. Unlike in the blocked checkerboard antiferromagnetic order, here half of pairs of the next nearest neighbor Ti moments bridged by O atoms are in parallel while the other half are in anti-parallel, likewise with Pn atoms.

For compounds Na$_2$Ti$_2$Pn$_2$O the moment around each Ti atom is found to be about $0.5\mu_B$ in both blocked checkerboard and bi-collinear antiferromagnetic orders. This is consistent with the spin density wave mechanism, in which the moment should be small since its formation is only attributed to those states nearby the Fermi energy. Considering there is a large amount of antiferromagnetic quantum fluctuation in low dimensional systems, the effective ordering moment around each Ti atom will be further much reduced. This is likely the reason why no magnetic ordering in compounds Na$_2$Ti$_2$Pn$_2$O has so far been detected by the powder neutron diffraction.\cite{bao}

A bi-collinear antiferromagnetic order was first introduced in the study of Fe-based superconductors to describe the ground state of $\beta$-FeTe,\cite{ma3} very interestingly as a strong evidence to against the view of spin density waves in the Fe-based superconducting parent compounds, since there is no any Fermi surface nesting vector to correspond to such a magnetic order, moreover there is a large magnetic moment of about 2.5$\mu_B$ formed on each Fe atom.

The substantial difference between the two cases lies in the fact that there is only one electron in 3$d$-orbital for a Ti$^{3+}$ ion, in contrast there are five electrons in 3$d$-orbital for an Fe$^{2+}$ ion. Accordingly it is not expected that there is strong electronic correlation effect in Na$_2$Ti$_2$Pn$_2$O. We have performed a GGA+$U$ calculation. We find that when $U$ is larger than 2~eV, the ferromagnetic state becomes the lowest energy state for both compounds. This contradicts the experimental observation that there is an antiferromagnetic-like transition around 300 K for Na$_2$Ti$_2$As$_2$O and around 120 K for Na$_2$Ti$_2$Sb$_2$O.\cite{ozawa2,ozawa3,ozawa4}

\subsubsection{Na$_2$Ti$_2$As$_2$O}

As aforementioned, the calculations show that any magnetic state, except the blocked checkerboard and bi-collinear antiferromagnetic states, is unstable and will relax into a nonmagnetic state for compound Na$_2$Ti$_2$As$_2$O.

\begin{figure}
\centering
\includegraphics[width=7.5cm]{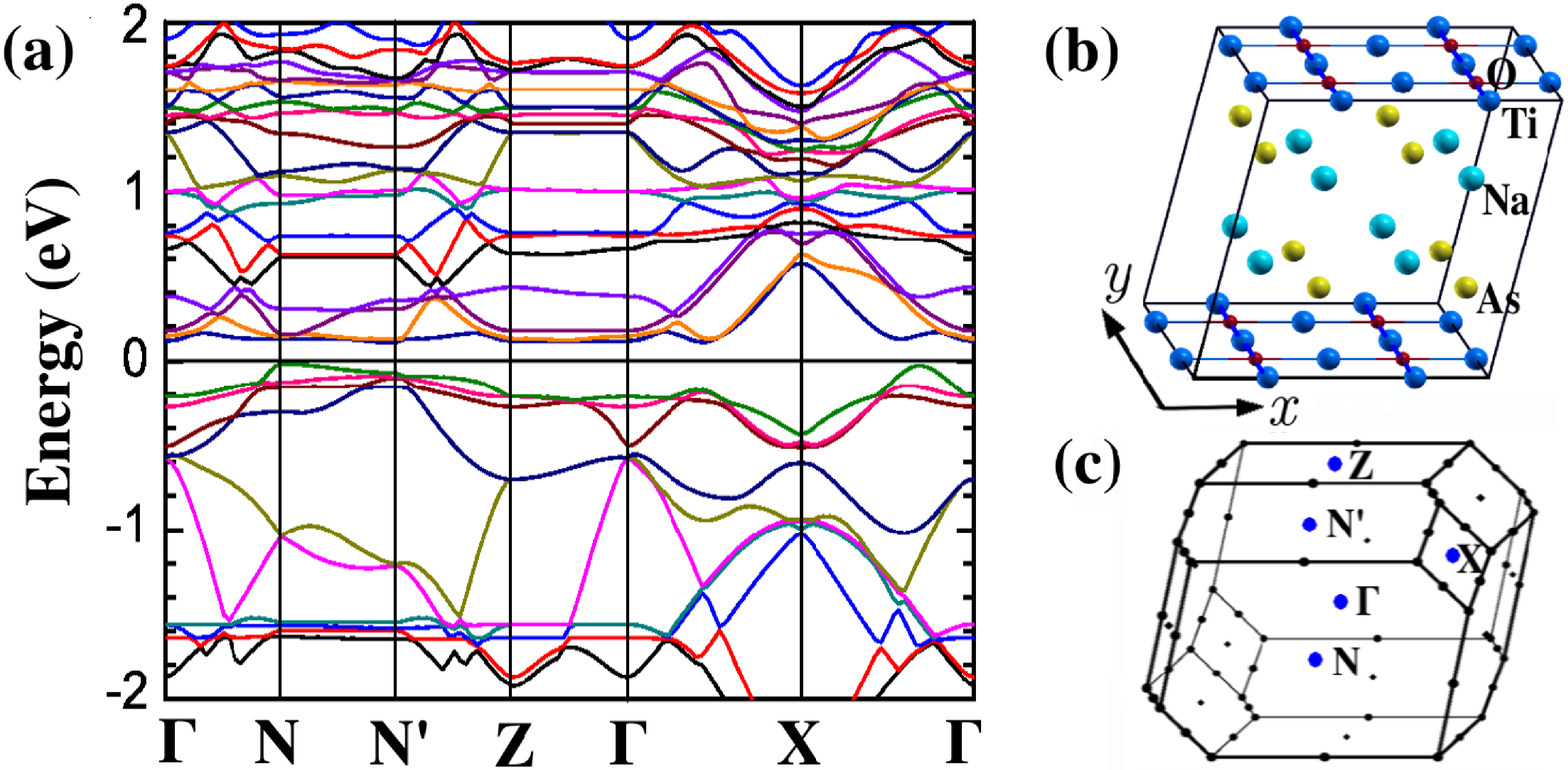}
\caption{(Color online) Na$_2$Ti$_2$As$_2$O in the blocked checkerboard antiferromagnetic state: (a) Band structure; (b) Magnetic unit cell; (c) Brillouin Zone with selected high symmetric points. Here the top of the valence band sets to zero.}\label{fig9}
\end{figure}

\begin{figure}
\includegraphics[width=7.5cm]{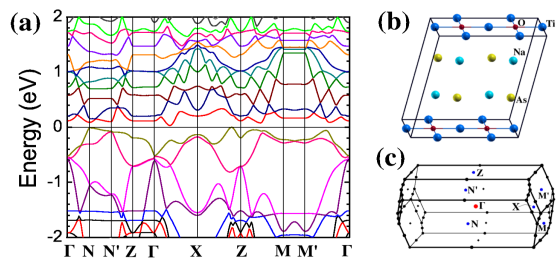}
\caption{(Color online) Na$_2$Ti$_2$As$_2$O in the bi-collinear antiferromagnetic state: (a) Band structure; (b) Magnetic unit cell; (c) Brillouin Zone with selected high symmetric points. Here the top of the valence band sets to zero.}\label{fig11}
\end{figure}

To calculate the blocked checkerboard antiferromagnetic (AFM) order in Na$_2$Ti$_2$As$_2$O, we construct such a crystal unit cell consisting of four formula cells, which is triclinic with three base vectors of (2$a$, 0, 0), (0, 2$a$, 0), and (0.5$a$, 0.5$a$, 0.5$c$) respectively, as shown in Fig. \ref{fig9}(b). For the bi-collinear antiferromagnetic order, the crystal unit cell is also constructed as a triclinic cell but with three base vectors of (2$a$, 0, 0), (0, $a$, 0), and (0.5$a$, 0.5$a$, 0.5$c$) respectively (see Fig. \ref{fig11}(b)).

After the full structural optimization, we find that compound Na$_2$Ti$_2$As$_2$O in the blocked checkerboard AFM order is energetically lower by about 4.5 meV/Ti than in the bi-collinear AFM order. Figures \ref{fig9} and \ref{fig11} show the electronic band structures of compound Na$_2$Ti$_2$As$_2$O in the blocked checkerboard AFM and bi-collinear AFM orders respectively. In both AFM orders, Na$_2$Ti$_2$As$_2$O is transformed into a semiconductor from a semimetal, with a small band gap of about 0.15 and 0.05 eV respectively; moreover the moment around each Ti atom is found to be 0.56 and 0.53 $\mu_B$ respectively. These calculated results are summarized in Table \ref{table1}.

\begin{table} \renewcommand{\arraystretch}{1.3}
\addtolength{\tabcolsep}{+1pt}
\caption{Calculated energies and magnetic moments of the nonmagnetic, ferromagnetic, checkerboard, collinear, bi-collinear, and blocked checkerboard antiferromagnetic states respectively, for Na$_2$Ti$_2$As$_2$O and Na$_2$Ti$_2$Sb$_2$O. The energy in the nonmagnetic state is set to zero. The bar ``--" in table means that the corresponding magnetic state is not stable and will relax into a nonmagnetic state.}
\begin{tabular}{l c c c c}
\hline
\hline
Magnetic order  & \multicolumn{2}{c}{Na$_2$Ti$_2$As$_2$O}
& \multicolumn{2}{c}{Na$_2$Ti$_2$Sb$_2$O} \\
\hline
&energy & moment & energy&moment \\
               & (meV/Ti) & ($\mu_B$) & (meV/Ti) & ($\mu_B$) \\
\hline
nonmagnetic             & 0    &  0    &  0    & 0  \\
ferromagnetic            & --   &  --    & --    & -- \\
checkerboard           & --   &  --    & --    &  -- \\
collinear      & --   &  --    & --    &  -- \\
bi-collinear   & -20.8& 0.53   & -19.5 &  0.53 \\
blocked checkerboard   & -25.3& 0.56   & -11.7 &  0.50 \\
\hline
\hline
\end{tabular}
\label{table1}
\end{table}

Table \ref{table2} lists the calculated lattice parameters with the selected bond lengths. There is a structural distortion found for compound Na$_2$Ti$_2$As$_2$O in the blocked checkerboard AFM state, in which the four Ti atoms in a magnetic block slightly gather towards their center, an As atom. This gathering leads to the Ti-O bond slightly elongating. Correspondingly, the bond distances among the four Ti atoms in a magnetic block are reduced. These four Ti atoms will form a compact square. As a result, the Ti-Ti and Ti-As chemical bonds within each magnetic block become stronger. Such a tetramer lattice distortion is certainly due to the spin-lattice coupling since it cannot happen in any nonmagnetic state. And it can be quantitatively described by the bond distance ratio of O-Ti-O/As-Ti-As, as reported in Table \ref{table2}. Such a distortion is in good agreement with the powder neutron diffraction observation that there is a commensurate structural distortion but the space group symmetry is retained after the anomaly transition.\cite{ozawa3}

\begin{table}
\caption{\label{table2} Calculated lattice parameters with some selected bond lengths in the blocked checkerboard antiferromagnetic state versus nonmagnetic state for Na$_2$Ti$_2$As$_2$O. }
\begin{tabular}{l c c}
\hline
\hline
                & nonmagnetic & blocked checkerboard \\
\hline
a                     &  4.0597  & 4.0701  \\
c                     & 15.389 & 15.364   \\
Ti-O                  & 2.0299 & 2.0347/2.0352 \\
Ti-As                & 2.7214 & 2.7175/2.7281  \\
O-Ti-O/(As-Ti-As)   & 0.7459 & 0.7474 \\
\hline
\hline
\end{tabular}
\end{table}

There is also a structural distortion found for compound Na$_2$Ti$_2$As$_2$O in the metastable bi-collinear AFM state, which will be discussed in detail below in association with compound Na$_2$Ti$_2$Sb$_2$O.

\subsubsection{Na$_2$Ti$_2$Sb$_2$O}

\begin{figure}
\includegraphics[width=7.5cm]{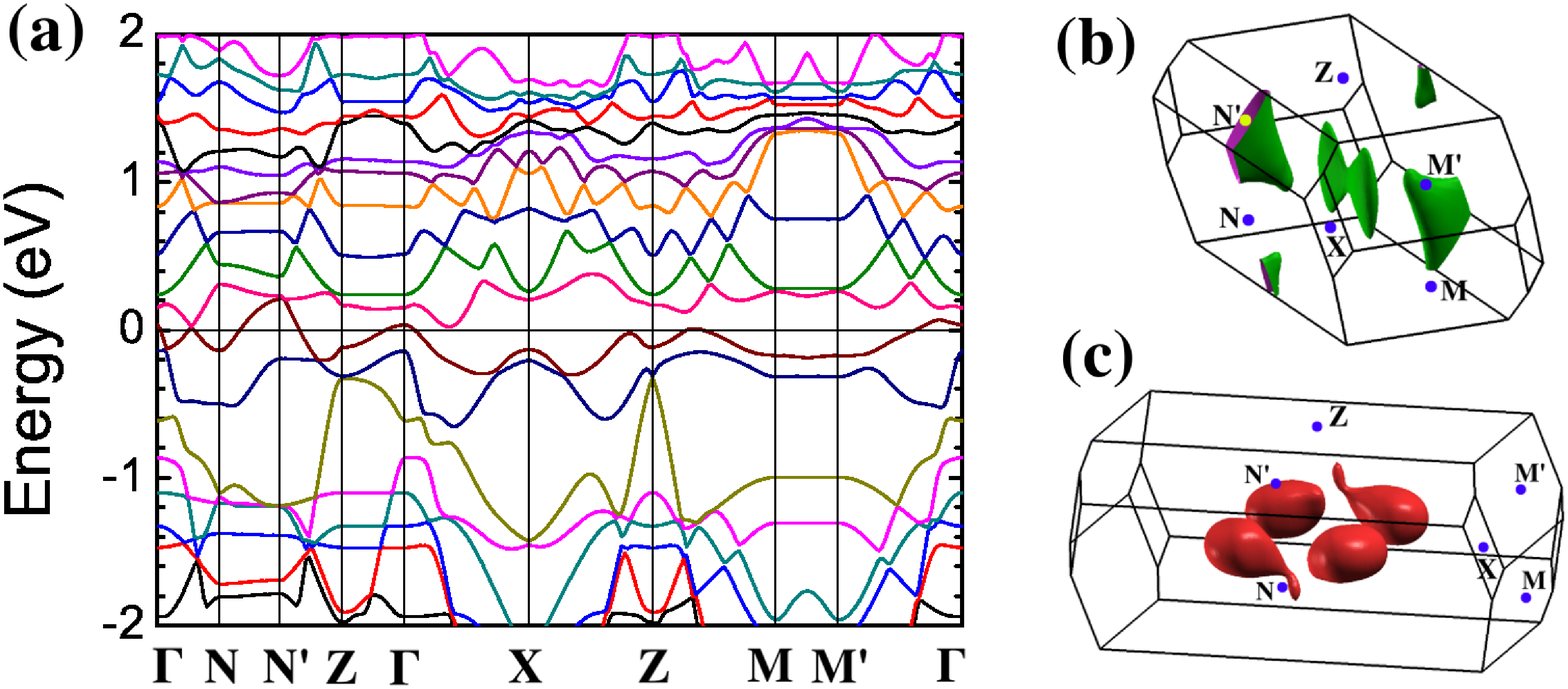}
\caption{(Color online) Na$_2$Ti$_2$Sb$_2$O in the bi-collinear antiferromagnetic state: (a) Band structure; (b) Hole-type Fermi surface in the Brillouin Zone; (c) Electron-type Fermi surface. The Fermi energy is set to zero.}\label{fig12}
\end{figure}

\begin{figure}
\centering
\includegraphics[width=7.5cm]{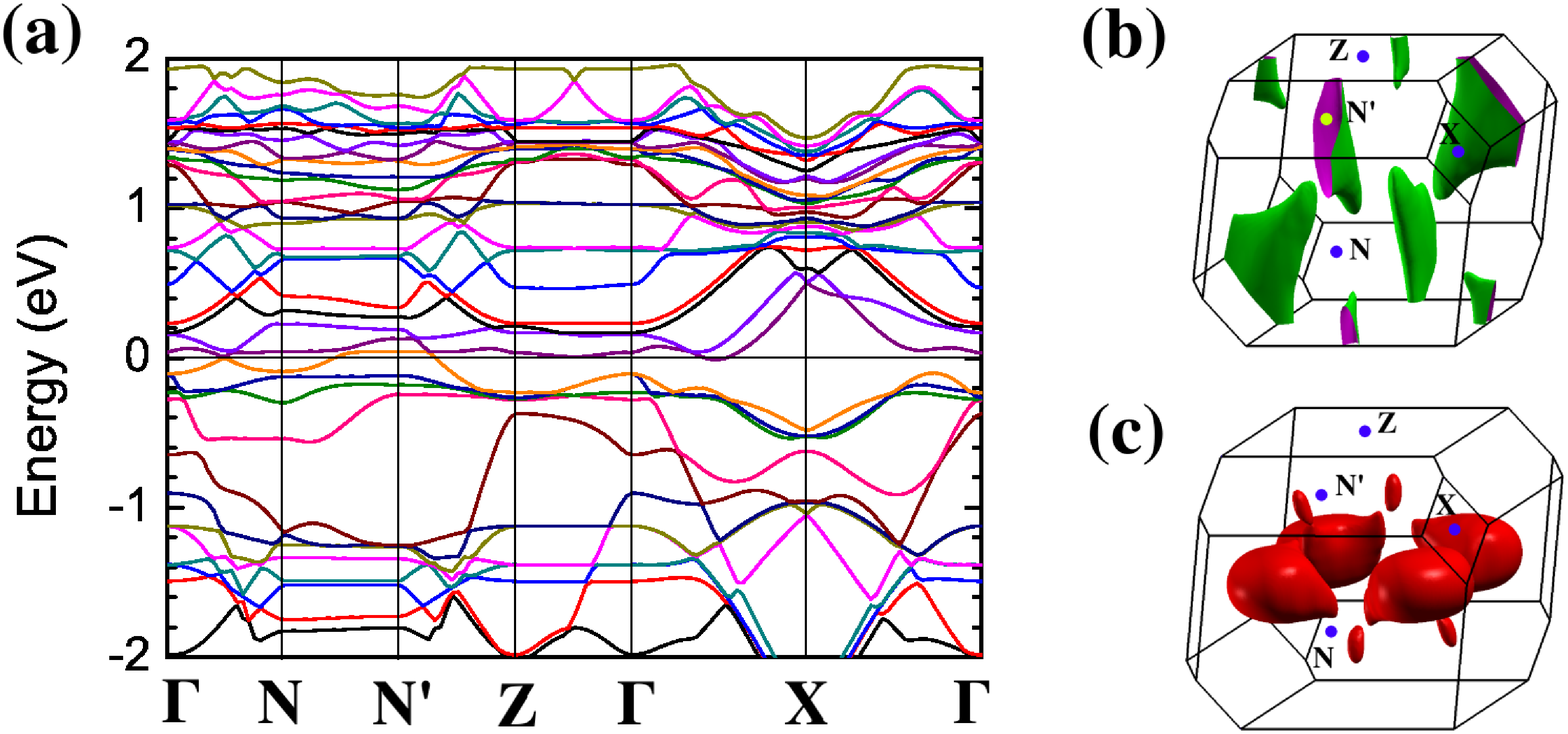}
\caption{(Color online) Na$_2$Ti$_2$Sb$_2$O in the blocked checkerboard antiferromagnetic state: (a) Band structure; (b) Hole-type Fermi surface in the Brillouin Zone; (c) Electron-type Fermi surface in the Brillouin Zone. The Fermi energy is set to zero.}\label{fig10}
\end{figure}

We performed the similar calculations on compound Na$_2$Ti$_2$Sb$_2$O. However, we find that compound Na$_2$Ti$_2$Sb$_2$O in the bi-collinear AFM order is the lowest in energy, about 8 meV/Ti lower than in the blocked checkerboard AFM order. In Figs. \ref{fig12} and \ref{fig10}, we respectively show the calculated electronic band structures of Na$_2$Ti$_2$Sb$_2$O in the bi-collinear and blocked checkerboard AFM orders. In Fig. \ref{fig12} (b), the high symmetry directions $N'-Z$ and $M'-Z$ correspond to the moment parallel and anti-parallel alignments respectively. The shapes of the Fermi surface along these two directions are distinct. The volume enclosed by the electron-type Fermi surface sheets is larger than that enclosed by the hole-type Fermi surface sheets, which means that the electron carrier density is larger than the hole carrier density. It is noticed that the shapes of Fermi surface in both AFM orders are quite similar.

Unlike Na$_2$Ti$_2$As$_2$O, Na$_2$Ti$_2$Sb$_2$O remains semi-metallic in both AFM orders. This is consistent with the experimental observation that the electric resistivity of Na$_2$Ti$_2$As$_2$O is higher than that of Na$_2$Ti$_2$Sb$_2$O by approximately a factor of 10 after the anomaly happening.\cite{ozawa3} Examining the Fermi surfaces of Na$_2$Ti$_2$Pn$_2$O in the nonmagnetic state (Fig. \ref{fige}), we see that Na$_2$Ti$_2$As$_2$O has nearly perfect two-dimensional isotropic nesting between the Fermi surface sheets centered around X and M, which is expected to induce a full energy gap opened at the Fermi energy after the spin density wave transition. In contrast, Na$_2$Ti$_2$Sb$_2$O shows obvious anisotropic nesting between the Fermi surface sheets centered around X and M. Such an anisotropic nesting can only induce a partial energy gap opened at the Fermi energy. This is the underlying physics that Na$_2$Ti$_2$As$_2$O becomes semiconducting while Na$_2$Ti$_2$Sb$_2$O remains semi-metallic after the spin density wave transition, and may explain why the former in the isotropic blocked checkerboard AFM order is more favorable energetically while the latter in the anisotropic bi-collinear AFM order.

Table \ref{table3} summarizes the calculated lattice parameters with selected bond lengths. The calculations show that there is also a small structural distortion for compound Na$_2$Ti$_2$Sb$_2$O in the bi-collinear AFM state, similar to the case of $\beta$-FeTe in the bi-collinear AFM state. The lattice constant becomes slightly longer along the spin anti-parallel alignment to lower AFM energy and shorter along the spin-parallel alignment to lower further ferromagnetic energy. As a result, the crystal primitive unit cell on Ti$_2$Sb$_2$O layer deforms from a square to a rectangle.

Compound Na$_2$Ti$_2$Sb$_2$O in the metastable blocked checkerboard AFM state shows a similar tetramer structural distortion as found in Na$_2$Ti$_2$As$_2$O above.

\begin{table}
\caption{\label{table3} Calculated lattice parameters with some selected bond lengths in the bi-collinear antiferromagnetic state versus nonmagnetic state for Na$_2$Ti$_2$Sb$_2$O. }
\begin{tabular}{l c c}
\hline
\hline
                & nonmagnetic & bi-collinear  \\
\hline
a (b)              & 4.1369  & 4.1059 (4.1851) \\
c                     & 16.6428 & 16.6166   \\
Ti-O                  & 2.0684 & 2.0486/2.0924  \\
Ti-Sb                 & 2.8928 & 2.8605/2.9059  \\
O-Ti-O/(Sb-Ti-Sb)   & 0.7150 & 0.7162/0.7201 \\
\hline
\hline
\end{tabular}
\end{table}

\section{Summary}

We have presented the first-principles calculations of the electronic structures of layered pnictide-oxide Na$_2$Ti$_2$Pn$_2$O (Pn=As, Sb). We find that the ground state of Na$_2$Ti$_2$As$_2$O is a blocked checkerboard antiferromagnetic semiconductor with a small gap of about 0.15 eV, while the ground state of Na$_2$Ti$_2$Sb$_2$O is a bi-collinear antiferromagnetic semimetal; and both have a small moment of about 0.5$\mu_B$ around each Ti atom. Moreover we find that there is a tetramer structural distortion in the blocked checkerboard antiferromagnetic state, in which the four Ti atoms at each magnetic block gather towards their center of an As atom, in good agreement with the experimental observation of a commensurate structural distortion but the space group symmetry retained after the anomaly happening. We confirm that there is a strong Fermi surface nesting in Na$_2$Ti$_2$Pn$_2$O. And we verify that the two antiferromagnetic states both are induced by the Fermi surface nesting rather than based on local magnetic moments. Actually these two antiferromagnetic states are close in energy for each of Na$_2$Ti$_2$Pn$_2$O, in which one is the ground state, then the other is the metastable state. We may consider layered pnictide-oxides Na$_2$Ti$_2$Pn$_2$O as a paradigm for spin density waves.

\begin{acknowledgements}

We wish to thank Professor Xian-Hui Chen for bringing the layered pnictide-oxides to our notice. This work is supported by National Program for Basic Research of MOST of China (Grant No. 2011CBA00112) and National Natural Science Foundation of China (Grant Nos. 11190024 and 91121008).

\end{acknowledgements}

{\it Note added.} While in the preparation of this manuscript, we learnt of a paper by David Singh (arxiv1209.4668), which reports that the instability at $X$ point associated with the Fermi surface nesting leads to a double striped antiferromagnetic (namely bi-collinear AFM) state in compound BaTi$_2$Sb$_2$O. We have subsequently examined a variety of magnetic states for compounds BaTi$_2$Pn$_2$O (Pn=As, Sb) and find that none of them is stable except the blocked checkerboard and bi-collinear antiferromagnetic states, which are however nearly degenerated with the nonmagnetic state with a lower energy of no more than 0.5 meV/Ti and the moment around each Ti atom is just about 0.2~$\mu_B$. The Fermi surfaces that we obtained for BaTi$_2$Pn$_2$O are similar to those shown in Figs. 6 and 7 reported in the arxiv1209.4668 paper, in which the Fermi surface nesting clearly is dramatically weaker than those in compound Na$_2$Ti$_2$Pn$_2$O reported in the present paper (see Fig. \ref{fige}).



\begin{thebibliography}{99}
\bibitem{bao}The powder neutron diffraction resolution upon an ordering moment is about $0.5\mu_B$, by private commnucations with Professor Wei Bao.
\end{thebibliography}
\end{document}